# Software-intensive product engineering in start-ups: a taxonomy

*E. Klotins, M. Unterkalmsteiner, T. Gorschek*

Software Engineering Research Lab Sweden
Blekinge Institute of Technology
371 79 Karlskrona
Sweden

Software start-ups are new companies aiming to launch an innovative product to mass markets fast with minimal resources. However a majority of start-ups fail before realizing their potential. Poor software engineering, among other factors, could be a significant contributor to the challenges experienced by start-ups.

Very little is known about the engineering context in start-up companies. On the surface, start-ups are characterized by uncertainty, high risk and minimal resources. However, such characterization is not granular enough to support identification of specific engineering challenges and to devise start-up specific engineering practices.

The first step towards understanding on software engineering in start-ups is definition of the Start-up Context Map - a taxonomy of engineering practices, environment factors and goals influencing the engineering process. Goal of the Start-up Context Map is to support further research on the field and to serve as an engineering decision support tool for start-ups.

## 1   Introduction

Software start-ups are small companies focusing on engineering and launching innovative products fast. Such companies are gaining momentum as important part of the economy and are central for innovation [1]. However, the failure rate among start-up companies is high. Various sources suggest that 75-95% of start-ups fail before realizing their potential [2], [3]. In 2015 alone, more than $429 billion were invested in start-ups and given an optimistic 75% failure rate, that amounts to $322 billion wasted on building unsuccessful products [4], [5]. High failure rates could be explained with unfortunate market conditions, flaws in a business model, lack of commitment or, simply put, a bad product idea. However, the impact of inadequate software engineering practices to the failure rate is largely unknown and could be a significant contributing factor next to marketing and business issues [6], [7].

Very little is known about the engineering in start-ups. On the surface, start-ups are characterized by uncertainty, lack of resources and high risk [8], with product engineering and marketing being the top challenges [6]. However, this characterization is not granular enough to identify specific challenges or to support adaptation of engineering practices from other, similar contexts, and subsequently validate start-up specific engineering practices.

Some engineering challenges in start-ups can potentially be solved by adapting scaled down practices from established companies [7]. However, lack of understanding about the engineering context makes adaptation and validation of practices in start-ups difficult [9]. For example, scaled down agile practices that attempt to address team collaboration issues, which might be negligible in a three-person start-up. As a result, ad-hoc adaptation of engineering practices depends on the experience of lead engineers, but can produce waste, and often delivers unanticipated results [10], [11].

The first step towards start-up specific engineering practices is to understand the engineering context of start-ups. This enables adaptation and development of start-up specific engineering practices. A context can be described, for example, by listing factors influencing the engineering process, and breaking them down [12]. However, simply listing context factors is not enough. The context description must be presented in a way that is useful for decision support in practice to be used by practitioners, but also enable further research on engineering in start-ups.

In this paper we present the Start-up Context Map, a breakdown and description of factors influencing software-intensive product engineering in start-ups. Through detailed descriptions of factors influencing engineering processes the Start-up Context Map is also a decision

> With software-intensive product engineering we understand application of well-understood practices in an organized way to evolve a product, containing a non-trivial software component, from idea to market, within cost, time and other constraints.
>
> The engineering process is influenced by a set of circumstances, i.e. context. Understanding of context is a crucial part of understanding the engineering process [15].

support tool, and a repository, containing state-of-the-art and state-of-the-practice knowledge. Along with the context map we present four use scenarios tailored for both practitioners and researchers. Notably, the start-up context map is the first attempt to systematize knowledge pertaining software-intensive product development in start-ups.

Engineers and entrepreneurs can use the Start-up Context Map mainly as an engineering decision-making support tool, but also to learn from others and share their experiences with peers. Equally important, researchers can use the map to facilitate transfer of their research results to practice, and get input from practitioners. In essence the Start-up Context Map can be seen as a Software Engineering Body of Knowledge for start-ups that is open, evolving as new experience is gained, and managed by the start-up community. While the map is at this time still in a draft state, there are currently no alternatives neither for practitioners, nor researchers, to provide decision support and to devise better engineering practices for start-ups [13].

## 2  Evolution of the map

We use the industry-academia collaboration model proposed by Gorschek et al. [14] as a research framework (Figure 1). The model to illustrates the evolutionary steps of the Start-up Context Map from inception to an envisioned future. White bubbles represent completed steps; grayed bubbles show envisioned future evolution of the map.

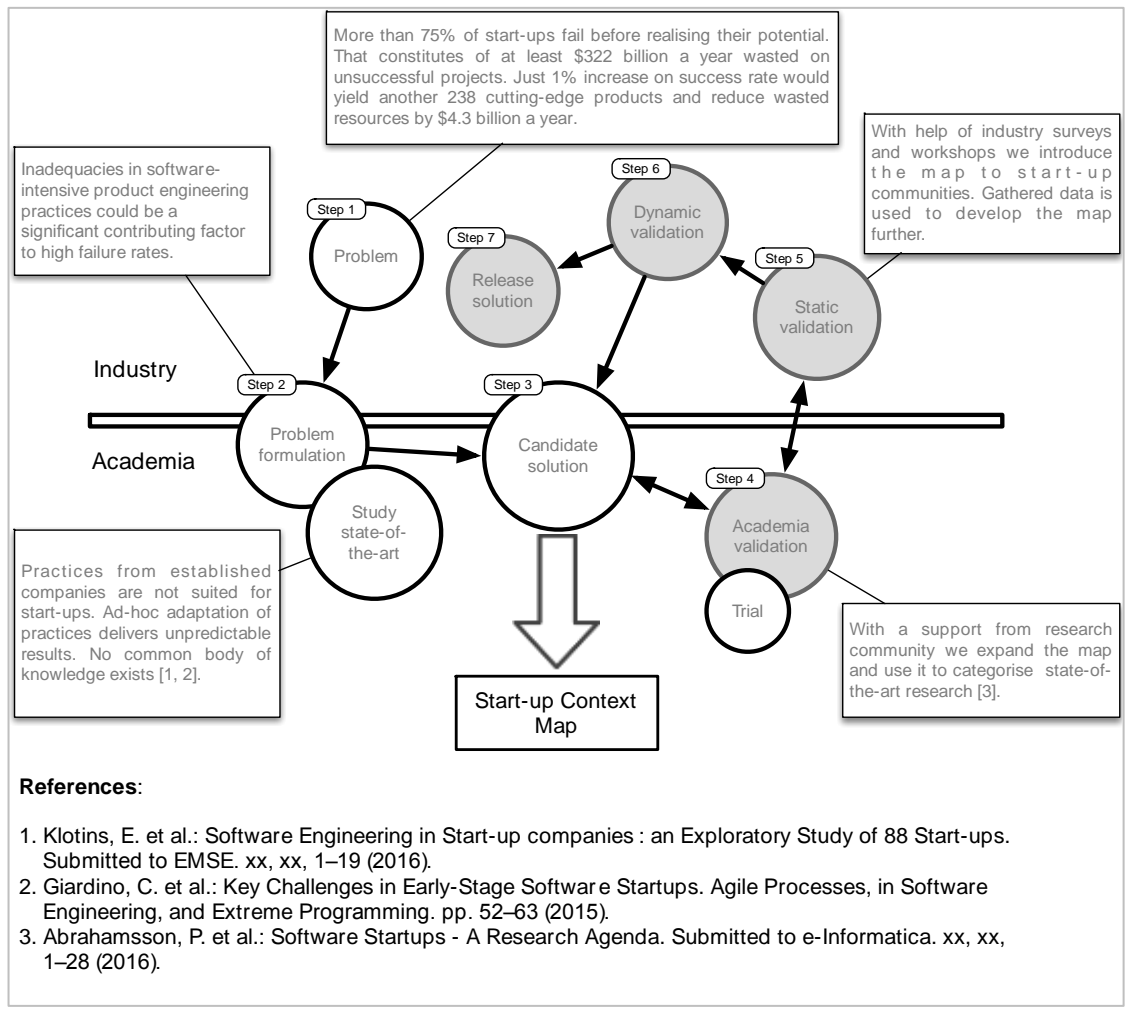

Figure 1, evolution of the Start-up Context Map, adapted from Gorschek et al [14].

The high failure rate of start-ups poses an economical problem for the software industry (Step 1, Figure 1). Based on literature surveys and the study of start-up experience reports, we formulated the working hypothesis that inadequacies in engineering could be a contributing factor to these high failure rates (Step 2). While start-ups adapt practices from established companies, such practices are not necessarily suited for start-ups as they attempt to address challenges not present in start-ups and disregard start-up specific challenges [6], [7]. This finding highlights the need to identify and/or develop, evaluate and validate more start-up specific software engineering practices.

As a first contribution towards start-up specific engineering practices, we present the Start-up Context Map (Step 3). The aim of the map is to characterize engineering context in start-ups, thus supporting decision-making in start-ups and providing a foundation to develop start-up specific engineering practices. We used other context models as an inspiration for parts of the map. The construction and use of the Start-up Context Map is presented and discussed in detail in Section 3.

The map is a continuous work in progress. For this reason we have created an online tool (http://startupcontextmap.org) supporting collaboration and continuous development of the map. The functionality of the tool and its usage scenarios are described in Section 3.3 and illustrated in Figure 2.

We conducted a validation of the map by linking 88 start-up experience reports to specific nodes in the Start-up Context Map (Step 4). This can be seen as a trial of the map showing usefulness of the map to systematize unstructured information as well as validating the contents and coverage of the map. As exemplified in 3, statements from the experience reports complement the map with first-hand practitioner experience on specific factors.

To complete Step 4 we aim to conduct workshops within the start-up research community[1]. Insights gained from the workshops will be used to update the map and prepare it for static and dynamic industry validation, shown as Steps 5 and 6 in Figure 1. We expect to receive specific and actionable input from the workshops thus at this point it is difficult to provide a design of subsequent validation steps. However, our overall strategy is to seed the Start-up Context

---

**Research method A: Snowball sampling**

Snowball sampling is a reference-based method to systematically discover new papers referenced by a starting set [1]. We applied the method in 4 steps:

**Definition of a start set:** We started by a paper by Petersen et al [2] and looked for other similar papers aiming to categorize context. We consulted colleagues and looked into different research areas for papers aiming to characterize context. The final set consisted of 5 papers from start-up [4], agile [5], SPI in small organizations [3] and industrial context [2] research areas.

**Discovery of papers:** We performed multiple iterations of forward snowball sampling until no new relevant papers were discovered.

**Screening**: The discovered papers were screened to remove duplicates, not peer reviewed sources and papers that do not contain a listing of engineering context factors.

**Sanity check:** We performed a backward snowball sampling iteration to make sure we have not missed any relevant paper.

**References**:

1. Wohlin, C.: Guidelines for Snowballing in Systematic Literature Studies and a Replication in Software Engineering, EASE 2014
2. Petersen, K., Wohlin, C.: Context in industrial software engineering research. 2009 3rd Int. Symp. Empir. Softw. Eng. Meas. 401–404 (2009).
3. Dybaa, T.: Factors of Software Process Improvement Success in Small and Large Organizations: An Empirical Study in the Scandinavian Context. Proc. 9th Eur. Softw. Eng. Conf. Held Jointly with 11th ACM SIGSOFT Int. Symp. Found. Softw. Eng. 7465, 148–157 (2003).
4. Chorev, S., Anderson, A.R.: Success in Israeli high-tech start-ups; Critical factors and process. Technovation. 26, 2, 162–174 (2006).
5. Chow, T., Cao, D.-B.: A survey study of critical success factors in agile software projects. J. Syst. Softw. 81, 6, 961–971 (2008).

---

[1] For example, Software Start-up Research Network, http://startupresearchnetwork.org

Map such that it provides immediate and growing value for practitioners, stimulating the sharing of data that can be synthesized by researchers.

## 3 The Start-up Context Map

### 3.1 Construction of the map

We conducted a literature review with the goal to find existing categorizations of goals, practices and environment factors from other contexts (see the sidebar on Research Method A).. From the 1206 discovered papers, we identified 9 studies proposing a categorization of engineering context. However, none of the existing models were broad enough, to cover the envisioned scope of the context map – start-up goals, environment and practices. The existing models are overlapping and even sometimes incompatible with each other.

Lacking a solid starting point, we created a structure of the map by brainstorming, plugging-in existing models in our structure and searching for additional studies supporting each node in the Start-up Context Map. During the construction process we continuously cross-validated the map with existing models to assure that all relevant context factors from existing models are placed in the Start-up Context Map.

### 3.2 Structure of the map

The map is a collection of factors relevant for software engineering in start-ups. It has three main factors, practices, categorizing engineering practices, goals, categorizing goals that are relevant for software engineering, and environment, categorizing environment surrounding the engineering process. Practices, goals and environment affect each other and must be considered together.

Any decisions in a start-up are driven by goals, thus the characterization of goals is the first, out of three, main context factors. Decisions are made in a specific environment that influences the decisions, thus the characterization of the environment is the second main context factor. Decisions pertain utilization of specific practices in a specific environment. Thus, the characterization of practices is the third main context factor. The top panel of Figure 2 shows the overview of the context map. Engineering practices, environment and goals are the main context factors and are broken down into sub-factors.

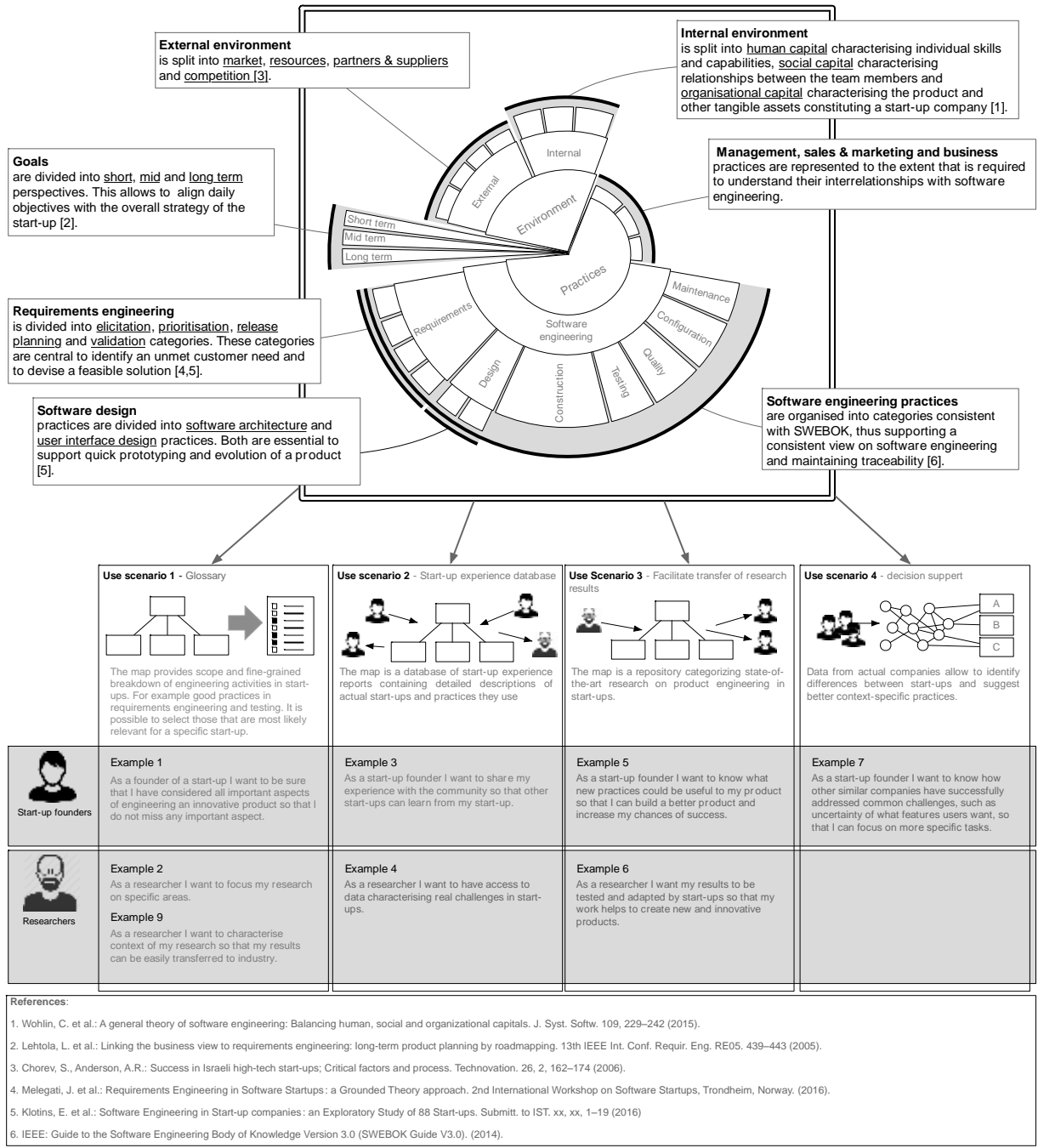

Figure 2, overview of the Start-up Context Map and its use-scenarios

We have further detailed each main context factor by breaking it down into sub-context factors and leaf nodes creating a tree-like structure. At present the map holds 26 factors pertaining start-up goals, 77 factors pertaining internal and external environment, and 138 practices. Respective width of each segment denotes how many sub-factors it contains.

Each leaf node of the map contains a granular description of the factor, examples from practice, references to related work, and practitioners' experience (see Figure 3). Details supported by literature are referenced while details that still require validation are not. In this way we ensure a clear connection between the map and supporting literature.

Moreover, by denoting less certain details of the map we invite collaborators to investigate these details.

### 3.3 Validation of the map

We have performed several activities with an aim to validate the map. In particular, we have looked into completeness of the map and usefulness of the map to categorize information.

Completeness of the context map determines to what extent of the map covers all relevant aspects of software engineering in start-ups. Usefulness of the map determines to what extent structure of the map is suited for categorizing information regarding software engineering in start-ups.

#### 3.3.1 Assessing completeness of the map

One important concern is whether the scope the context map is sufficient to cover all relevant software engineering aspects of start-ups. We turned to a community of start-up researchers[2] and invited them to collaborate on validating the map. We conducted a joint workshop with a goal to review the map and to reach a consensus that it covers all relevant aspects of software engineering in start-ups. Participants of the workshop were first asked to review the map off-line and to provide suggestions for improvement. Then their feedback was compiled and unclear items discussed in an on-line meeting. A total of 10 researchers provided their input.

From the feedback we received we adjusted the scope of the map and updated its contents. For example, we added factors related to technical debt and improved description of factors related to software testing and construction. Besides and external helped to spot many language and formulation related issues.

The improved version of the map was used to devise a case survey for collecting data on engineering practices in start-ups (see use scenario 2).

**Lessons learned:**
- Scope of the Start-up Context Map is sufficient to describe software engineering in start-ups to a level that enables assessment of engineering context and utilized practices.
- There is a substantial interest from the research community to collaborate on the Start-up Context Map

#### 3.3.2 Assessing usefulness of the map

To test usefulness of the map we performed a static validation and applied the map on 88 start-up experience reports. The reports are written by one of the principals of a start-up. While such reports contain valuable experience lessons, their unstructured nature makes them difficult to benefit from.

We applied the context map to categorize statements in the experience reports. First, we segmented the reports by paragraph. Then, we analyzed each report paragraph by paragraph in order to identify statements pertaining specific engineering factors. Higher-level statements are mapped to higher-level nodes; more specific factors are mapped to leaf nodes of the map. Paragraphs addressing multiple factors or ambiguous statements are mapped to multiple nodes of the map.

---

[2] The Software Start-up Research Network, https://softwarestartups.org/}

In case we identified a statement that could not be clearly mapped to a factor, we considered expanding the map with a new factor.

As a result of the mapping, we linked 876 statements from the reports to 69 context factors. During the mapping process we identified and added new context factors to the map. For example, we now differentiate between internal and external view on a product. The internal view refers to an engineering view of a product, i.e. architectural design, effort, and complexity. The external view represents how a customer sees a product, for example, features, usefulness, and suitability for completing a task.

**Lessons learned from validation:**

- The Start-up Context Map is useful to systematize experience reports. Systematization enables topical access to statements in the reports, thus supporting further analysis. Descriptions of practitioners experience with a factor expand the description making the map more useful for practitioners.
- Practitioner experience reports are a useful source to identify new factors for the map, eventually also allowing pruning of the map if parts are never utilized.
- Using the map as an experience database is a feasible use scenario (see Figure 2).

### 3.4 Using the Context Map

The four bottom panels in 2 illustrate different use scenarios. The use scenarios are aimed to provide practical support to start-up practitioners and to connect start-ups with researchers, experts and other stakeholders.

**Use scenario 1** is to use the map as a glossary. A typical challenge for start-ups is to identify relevant focus areas and to avoid wasting resources on irrelevant activities. Moreover, a less experienced founder may not realize that a critical activity is neglected. As one start-up founder[3] reflected on the lack of understanding regarding the scope of a start-up: "*Looking back, when I started 99dresses fresh out of high school I was very naive and had zero idea what I was doing. In fact, I didn't even know what a startup was! I just knew I wanted to solve a problem I personally experienced: having a closet full of clothes but still nothing to wear.*" Even more experienced founders could benefit from the map, for example they could be biased towards their past experiences and overlook some context factors. A glossary therefore provides a light-weight decision-support tool that can be utilized by both novice and seasoned start-up practitioners.

The Start-up Context map provides a basis for creating a comprehensive glossary of good engineering practices in start-ups. Overview of the practices allow practitioners create their own methods. Use scenario 4 further elaborates identification of specific engineering methods. The glossary is community driven, entrepreneurs, start-up engineers and start-up researchers decide what practices are relevant to include in the map. By having an overview of what goals, environment factors and practices could be relevant, a founder can make an educated decision to focus on (or ignore, thereof) certain context factors.

---

[3] https://medium.com/female-founders/my-startup-failed-and-this-is-what-it-feels-like-c5d64b3ae96b, accessed July 8, 2016

Researchers can use the map to identify gaps and specific areas for investigation, e.g. by looking into less developed areas of the map. Equally important, researchers can use the map to characterize the context of their research, thus enabling safe adaptation of research results. By understanding the context around the application of a practice, successful transfer and replication of that practice can be enabled.

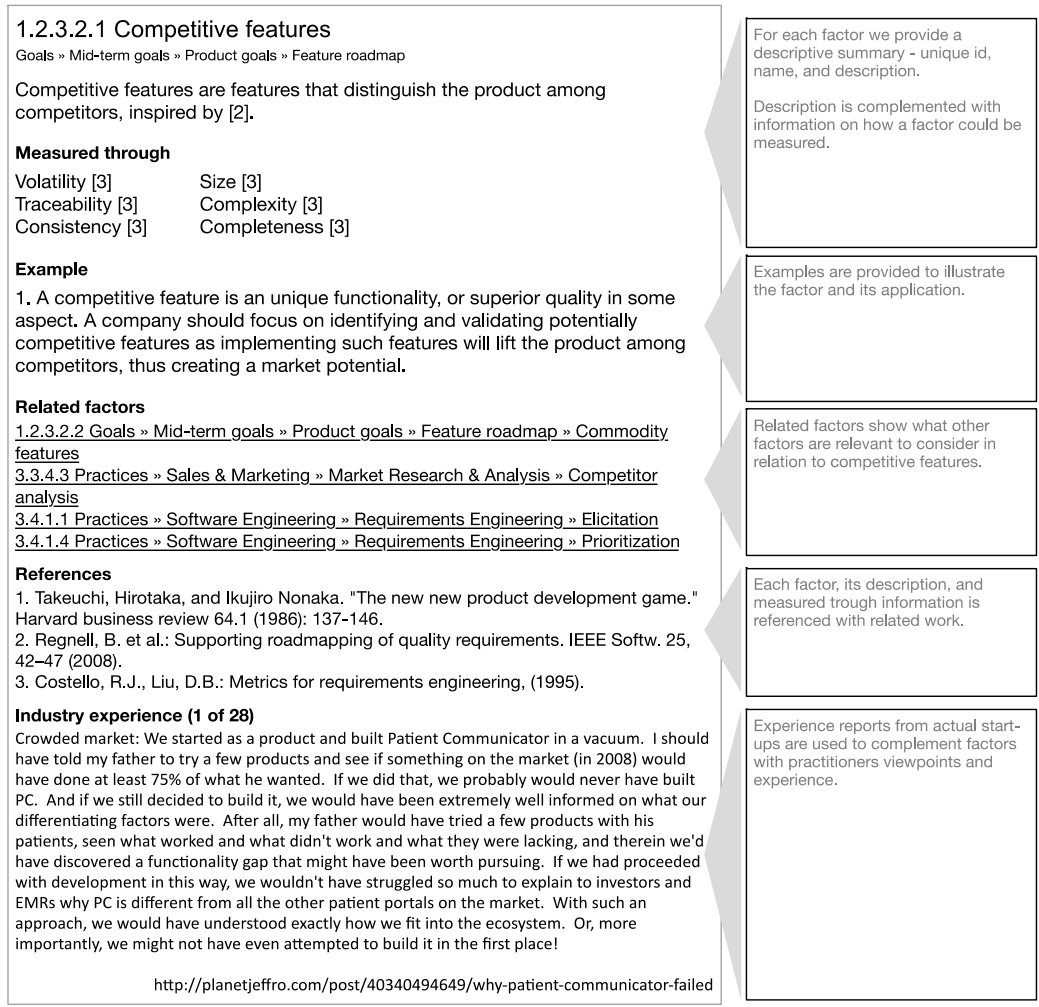

Figure 3, granularity of a leaf node

**Use scenario 2** refers to the utilization of the map as an experience database. While experience reports are valued both by practitioners and researchers, they often lack structure and essential details to be useful, not to mention that they are seldom collected in one place. As it stands now the map indirectly holds the combined experience of 88 start-ups, as well as being connected to research in the area.

Start-up founders are often keen to disseminate their experience. As one start-up founder states[4] on sharing his experience: *"One thing I'll be doing more of is writing about my experience. Partially because it's therapeutic, but also because if there's a silver lining in all of this (and there is), it's that I can help educate others about a path fraught with hardship, but rewarding nonetheless."*

---

[4] http://chrishateswriting.com/post/74083032842/today-my-startup-failed, accessed July 8, 2016

The Start-up Context Map provides a backbone for structured experience reports. Start-up engineers can add their experience with specific factors directly in the map (see Figure 3), thus enabling structured access to their report. In this way, the experience reports are easy to use by other practitioners as they can easily locate relevant parts of the report. Researchers can use the same structure to gather all experiences pertaining to a specific context factor and use that information as input for their research.

We have used the context map to devise a case survey[5] exploring how start-ups utilize engineering practices and what are characteristics of engineering context in start-ups. Results from the case survey will be used to update map with start-up engineering best practices.

**Use scenario 3** supports the transfer of research results from academia to start-ups. State-of-the-art practices from academia are created in a rigorous research process, has passed certain validation and technology transfer stages (see Fig.1), and is accompanied with sufficient description that allows to make an educated decision whether a practice is applicable or not in a specific start-up case.

Researchers can use the map and add their state-of-the-art results under specific nodes of the map. Therefore, start-up engineers can easily locate cutting-edge practices to address a specific issue they encounter. In combination with Use scenario 2 this enables a two-way collaboration between start-ups and researchers.

**Use scenario 4** describes the use of the map as decision support tool. The characterization of start-up context enables identification of similarities between start-ups, their goals, environments, and applied practices. The decision support system uses start-up environment characteristics and goals as input and suggests what engineering areas or environment factors should be in focus for a particular start-up. Inference engine and knowledge database for the decision support is based on start-up experience reports from use scenario 2. However, creating such decision support tool requires a substantial number of experience reports.

Thus, proven combinations of practices to tackle common challenges can be bundled together for reuse in other similar start-ups. Such bundles or "patterns" can be created by start-up experts and utilized by practitioners. The goal is to evolve patterns over time. Examples of such patterns are, for example, The Lean Start-up, XP, Scrum and various models aimed to support innovation in small organizations.

These use scenarios aim to reduce uncertainty by providing engineering decision support and by facilitating the development of new start-up context specific engineering practices. The Start-up Context Map, by enabling informed decisions, can therefore increase the likelihood of success in engineering innovative products in a start-up setting.

# 4 Conclusions

In this paper we present an attempt to improve software-intensive product engineering in start-ups. We propose the Start-up Context Map – characterizing start-up goals, internal and external environment, with a focus on engineering practices. The Start-up Context Map can be seen as start-up body of knowledge, similar to SWEBOK and PMBOK, but managed and developed by start-up community itself in collaboration with interested researchers.

---

[5] http://startupcontextmap.org/exp-survey/woifenw2

Our intention is to provide a core structure and necessary tools for the map to be further validated, adapted and evolved by the community. To support this the continuously evolving online tool [http://startupcontextmap.org](http://startupcontextmap.org) is launched.

## 4.1 Key points:

1. A structured approach to collecting start-up experience reports enables synthesis of "best engineering practices" for start-ups
2. Knowledge of what practices works and what's not under specific conditions is essential to tailor engineering process to suit needs of individual start-ups.
3. The Start-up Context Map is a community driven project to collect and categorize good start-up specific engineering practices.

## 5  About the authors

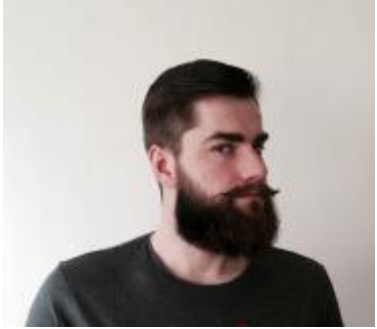
Eriks Klotins is a PhD student of Software Engineering at Blekinge Institute of Technology (BTH). The focus of his thesis is software engineering practices in start-ups. He has over nine years of experience in managing software development projects ranging from large government IT systems to several start-ups projects.
Contact him by: eriks.klotins@bth.se

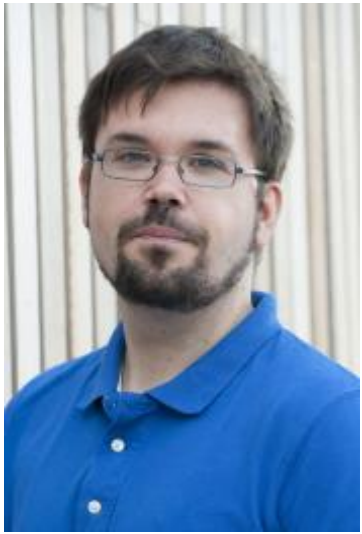
Michael Unterkalmsteiner received the BSc degree in applied computer science from Free University of Bozen-Bolzano in 2007, and the MSc and PhD degrees in software engineering from Blekinge Institute of Technology (BTH) in 2009 and 2015, respectively. He is a senior lecturer at BTH. His research interests include software repository mining, software measurement and testing, process improvement, and requirements engineering. He is a member of the IEEE. For more information or contact: www.lmsteiner.com, michael.unterkalmsteiner@bth.se

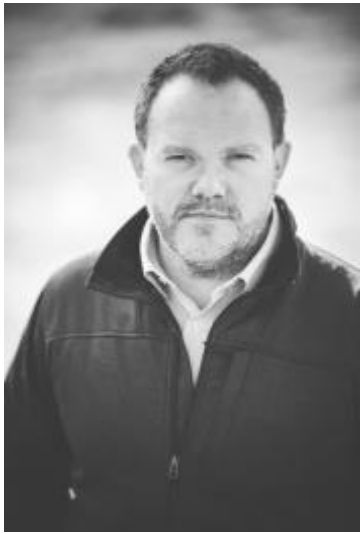
Tony Gorschek is a professor of Software Engineering at Blekinge Institute of Technology (BTH. He has over ten years' industrial experience as a CTO, senior executive consultant and engineer, but also as chief architect and product manager. In addition, he has built up five start-ups in fields ranging from logistics to internet based services.
Contact him by: tony.gorschek@bth.se